\documentclass[11pt,twoside]{article}
\usepackage{asp2004}
\usepackage{psfig}
\usepackage{epsf}
\usepackage{graphics}
\usepackage{lscape}
\begin{document}

\title{Chandra Studies of Millisecond Pulsars in Globular Clusters}
\author{Jonathan E. Grindlay}
\affil{Harvard-Smithsonian Center for Astrophysics, 60 Garden St.,
 Cambridge, MA 02138}

\begin{abstract}
The high resolution X-ray imaging and broad-band moderate resolution 
spectra enabled by the ACIS cameras on the Chandra X-ray Observatory 
have opened a new window in our study of millisecond pulsars (MSPs). Given 
the large excess of MSPs in globular clusters vs. the field, globular 
clusters are the favored sites for MSP study. Globular clusters (GCs) 
offer advantages of both known distance and likely MSP formation 
mechanisms but are complicated (or made more interesting!) by the 
complexities of subsequent MSP dynamical interactions. 
Here I review recent X-ray studies, focusing on our recent 
deep Chandra study of 47Tuc, which provide new constraints on 
both the spectra (thermal and non thermal) and evolution of MSPs and 
their encounters with both cluster and field stars.
\end{abstract}

\section{Introduction and Overview}
It is fitting that the oldest stellar systems in the Galaxy 
contain the oldest pulsars, the millisecond pulsars (MSPs), but not always  
in the oldest binary systems. Most MSPs in globular clusters (GCs) 
likely predate most of 
those in the field of the Galaxy given that the requisite neutron stars 
(NSs) in GCs were formed primarily from PopII massive stars 
which  predated NS production from massive stars in the disk. At 
the same time, some of the MSPs in GCs (e.g. PSR~B1821$-$24 
in M28) are clearly younger, as 
evidenced by their spindown ages $P/2{\dot{P}}$ (however approximate 
these may be). These suggest either recent spinup of old NSs, 
or more recent NS production from accretion induced collapse 
of white dwarfs in cluster binaries, or possibly {$\dot{P}$}~ values 
enhanced by mass transfer in a retrograde second exchange of an old MSP 
(see below). In fact it has become clear that some MSPs in GCs 
(e.g. 47TucW; discussed in more detail 
below) have had at least one swapped binary partner and thus 
more complex histories than any in the field. 

New clues into the nature, formation and evolution of MSPs have 
come from the additional constraints that $\sim$0.3--8\,keV 
X-ray observations of their integrated or pulsed emission provide. 
The high sensitivity, spectral resolution and bandwidth of the 
Chandra and XMM-Newton X-ray Observatories have enabled great 
progress in studies of MSPs. The great increase in 
angular resolution with Chandra has been essential for 
studies of MSPs in crowded GC cores. The first moderately deep Chandra 
observation of 47Tuc (Grindlay et al. 2001a; hereafter 
GHE01) provided a dramatic view of GC-MSPs and 
enabled (via X-ray to optical boresites) the   
optical identification, with HST, of the first MSPs in globulars. 
Accordingly, in this review we shall focus on Chandra (and some 
HST) results and only discuss briefly some recent XMM (and earlier ROSAT) 
studies.

The first optical identifications of MSPs in GCs, 
two in 47Tuc (Edmonds et al. 2001, 2002) and one in NGC 6397 
(Ferraro et al. 2001), have provided direct evidence 
that some GC-MSPs have swapped their binary 
partners. The main sequence or somewhat evolved binary companions 
now found in two such systems (PSR~J1740$-$5340 in NGC 6397 and 
MSP~W in 47Tuc, as discussed below) 
are not expected for systems that have evolved directly 
from their parent LMXBs. Post-LMXB secondaries are He white dwarfs 
(He-WDs) or very low mass degenerate remnants in field MSPs and 
most GC-MSPs. The nature and evolution of both of these systems 
is further constrained by their X-ray spectra and 
variability. The recently derived spectrum for MSP~W in 47Tuc 
(hereafter either 47TucW or MSP~W) shows 
hard emission, possibly from gas near the L1 point shocked by the 
pulsar wind (Bogdanov, Grindlay and van den Berg 2004, 
hereafter BGvdB04) and similar to that reported 
(Grindlay et al. 2001b, Grindlay et al. 2002; 
hereafter GCH02) for the Chandra source and MSP 
in NGC 6397, PSR~J1740$-$5340 (hereafter 6397A). 
Even more important is that the spectra, 
luminosity and X-ray vs. optical variability of 47TucW show 
remarkable similarities to the accreting MSP~J1808.4$-$3658, thus 
providing the crucial missing link between radio MSPs and LMXBs (BGvdB04). 

Recent Chandra studies of MSPs in GCs, as reviewed here, 
have now established that MSPs 
can show three different spectral components: very hard non-thermal emission 
(magnetospheric; with power law photon index $\sim$1) 
vs. hard (with PL index $\sim$1.5 and probably due to 
shocked gas) vs. soft (thermal emission from 
the NS polar caps, with typical kT $\sim$0.1\,keV). In 
this review we focus on the hard and soft components. 
Whereas both the very hard magnetospheric and soft 
thermal (polar cap) emission were evident from ROSAT spectra and timing 
(cf. Becker and Trumper 1999), the Chandra detection of the shocked gas 
component has required the high resolution imaging of Chandra to both 
isolate the source from neighbors and to enable HST identification of 
a heated stellar companion as the likely source of gas. It 
has also required sufficient temporal coverage to establish spectral 
variation with binary phase (for 47TucW).  Although pulsation spectra and 
pulsed light curves are also both highly desirable, spectra are arguably 
more important for a first investigation of MSP properties and have  
meant that the ACIS (with $\sim$150--200\,eV resolution), rather than the 
HRC (with virtually no energy resolution), Chandra cameras have been 
used in most GC-MSPs studies to date. Thus pulse-phased spectra 
for multiple MSPs imaged with Chandra are not possible (although this 
could be done in principle 
for a single MSP observed with the LETG grating and HRC-S), and HRC-S imaging 
pulsed light curves are only available for 
one MSP (J0437$-$4715, in the field; Zavlin et  al. 2002).  
As discussed by BGH04, such ``grey'' pulsed light curves can constrain 
MSP emission and neutron star properties.

\section{Brief Overview of X-ray Studies of MSPs in Globular Clusters}
\subsection{The Brightest and Hardest}
The X-ray brightest MSP in a globular cluster is the source in M28,
PSR~B1821$-$24, first detected and resolved 
with the ROSAT HRI (Danner et al. 1997) and 
for which high quality spectra (but with no temporal resolution 
for pulsations) have been obtained with Chandra (Becker et al. 2003). 
If this were the only MSP studied in X-rays, our view of their
properties would be very different: its predominantly non-thermal emission 
and narrow pulse profile in X-rays is very different from that now inferred  
(GCH02, Bogdanov et al. 2004, hereafter BGH04) 
for the bulk of the MSP population 
in 47Tuc with their predominantly thermal spectra (GCH02). 
Unambiguous X-ray spectra for even this apparently brightest GC-MSP require 
the spatial resolution of Chandra, given source confusion 
(Becker et al. 2003), although RXTE hard X-ray spectra for 
the (narrow) pulse peak (Kawai and Saito 1999) 
are in reasonable agreement. 

Upper limits have been obtained with RXTE for 
hard X-ray emission from the integrated population of MSPs in 
47Tuc (Ferguson et al. 1999), and RXTE has been able to provide 
spectra and timing constraints on the brightest nearby MSPs in the field. 
At $\gamma$-ray energies, the very 
flat non-thermal spectrum (with photon index $\sim$1) is detectable 
for B1821$-$24 and may have been detected with EGRET (cf. Zhang and Cheng
2003). MSPs in the field, which are generally much closer 
than the GC-MSPs, have been detected at $\gamma$-ray energies (e.g. 
J0218+4232, detected convincingly with EGRET by 
Kuiper et al. 1998). Its broad soft X-ray pulsed light curve suggests a 
thermal component, confirmed by its combined thermal plus 
PL spectrum detected with XMM-Newton (Webb, Olive \& Barret 2004).  
Detection of the non-thermal components of nearby thermally-dominated 
MSPs in the field (e.g. J0437$-$4715) may constrain the 
evolution of the magnetospheric spectra with the  
B field at the light cylinder, which may be correlated with 
non-thermal X-ray luminosity, as shown below.

\subsection{vs. the Dim and the Soft}
As suggested above, most MSPs in GCs are not bright and hard; B1821$-$24 is 
the exception. Rather, as our initial studies of the MSPs in 47Tuc 
have shown (GHE01, GCH02), the norm is relatively low X-ray luminosity, 
primarily in thermal emission and distributed in a relatively 
narrow band: $\sim$2--8$\times 10^{30}$ erg~s$^{-1}$. This was not apparent to 
ROSAT and the pioneering studies, and review, of Becker and Trumper
(1999) since again the samples were dominated by nearby or by extreme 
objects. The virtue of the GC-MSPs is, again, that they can define 
MSP parameters for entire populations in a single observation, given (of
course) the required angular resolution and sensitivity. 

This population-view, afforded by Chandra, can not only define the 
role of MSPs in binary evolution and thus cluster evolution, but in MSP 
or NS evolution as well. One of the key questions posed in GCH02 was: 
if MSPs are indeed swapping partners, as 
evidenced by  6397A or 47TucW (see 
below), then what must be the fate of i) the $P$ and {$\dot{P}$}~ history, if  
matter accretes with randomly aligned angular momentum vector 
${\bf J}$ from the new secondary onto the ``old'' MSP during 
the re-exchange, and ii) the $B$ field, both surface and at the light 
cylinder, if $B$ field evolution on NSs has anything to do with the 
accretion history of the star, as is commonly believed? We address 
some of these issues here and in more detail in forthcoming journal papers.

\subsection{From Hard to Soft (and back)?}
As shown below, the hard-bright MSPs have young ages ($P/2{\dot{P}}$) vs. the 
opposite for the dim-soft (reminiscent of other aging populations!). 
Does re-recycling restore youth? If so, could some of 
the apparently young MSPs (e.g. B1821$-$24) be 
masquerading youth after re-spinup in a second exchange, 
as the $B$ field is amplified by the differential rotation between 
a newly-accreted outer layer that is misaligned with the original ${\bf J}$ and ${\bf B}$
vectors? Or, more likely, is its NS a relic of 
the cluster formation and been able to maintain its stronger $B$ field 
without internally driven decay until its recent first-time accretion 
episode when it was exchanged in a binary 
encounter to acquire its first partner?

On the other hand, the soft-dim MSPs, like those which 
predominate in 47Tuc, 
may be resurrected (as for MSP~W) with a ``new'' main sequence 
companion apparently heated by the pulsar wind to produce mass loss 
through L1 that in turn 
is shocked by the pulsar wind to produce hard X-ray emission. This 
renewed hard emission phase for an MSP would be characterized by 
a lack of hard X-ray pulsations, though soft and sinusoidal 
pulsations are still expected from the polar caps. Resurrected MSPs  
would likely show a smooth 
modulation with binary phase if (as for 47TucW) 
part or all of the shocked-gas emission region is eclipsed by 
the companion. Such a doubly-recycled MSP can thus be hard but 
old (both chronologically and as measured by its new $P/2{\dot{P}}$). 
On the other hand, given the bizarre torques that must occur when 
a randomly aligned ``old'' MSP exchanges its (usually) degenerate 
secondary for a main sequence star and then has renewed mass transfer and 
spinup, the {\it deceleration} accretion torque from a retrograde 
second capture could cause the $P/2{\dot{P}}$ age to appear young 
despite a previously reduced B field.

These are representative questions raised by the fascinating 
demography of GC-MSPs. Let us look again at the MSP population in 47Tuc, 
but now with the additional perspective of a much deeper Chandra
observation. 

\section{Chandra Probes Deeper into 47Tuc}
Over a 13 day interval (September 29 -- October 11, 2002, 
we observed 47Tuc with 
four 65 ksec ACIS-S pointings and matched short (5ksec) sub-array 
exposures (to deal with bright sources). Simultaneous HST/ACS 
images were obtained in three filters (B, R and H$\alpha$) for 3 HST orbits 
on each of Chandra observations 2 -- 4. This Large Chandra 
Program was motivated in part by the MSP population, detected for 
the first time as soft X-ray sources with the less sensitive (at low 
energies) ACIS-I imager in our single 70ksec observation in 
March 2000 (GHE01). The overall source catalog for the ACIS-S observations 
is reported by Heinke et al. (2004a): some 300 sources are detected 
within the cluster half-mass radius of 2.8$\,^\prime$. In Figure 1 we 
show a representative color image constructed from the summed 
exposures in the 0.3--1.2\,keV (red), 1.2--2\,keV (green) and 
2--6\,keV (blue) bands. 

\begin{figure}
\plotone{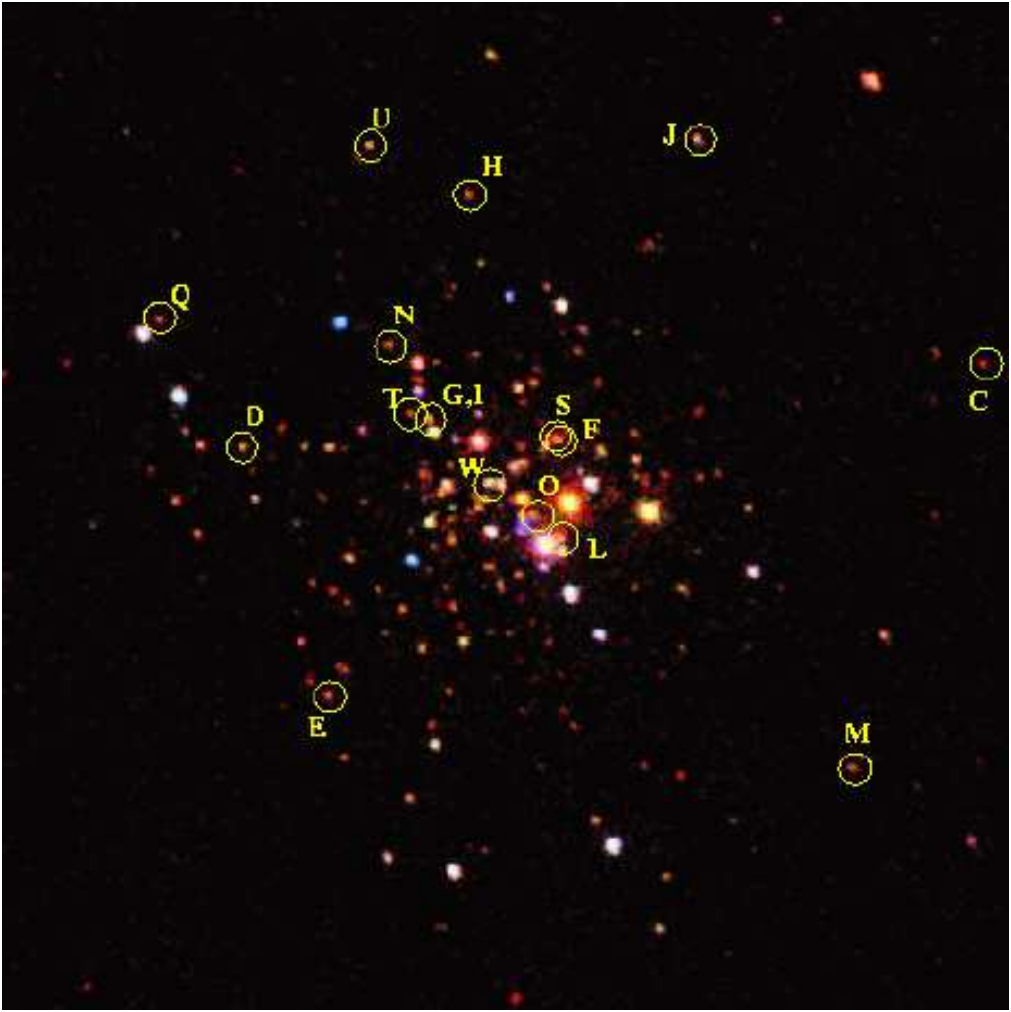}
\caption{Representative color image (from Heinke et al. 2004a) of 
central 2.5$\,^\prime$ $\times$ 2.5$\,^\prime$ of the deep ACIS-S exposure 
(270 ksec) on 47Tuc, with all 17 MSPs with known locations detected 
as marked. MSPs F and S, separated by 0.74$\,^{\prime\prime}$, are partially  
resolved but G and I, separated by 0.12$\,^{\prime\prime}$, are not. 
The Chandra positions are much more accurate than the 
large circles shown, with rms deviation from the MSP positions of 
only $\sim$0.2$\,^{\prime\prime}$. (NOTE: this image compressed for 
astro-ph size limits; original to appear elsewhere.)}
\end{figure}

All 16 MSPs located with timing positions 
(Freire et al. 2001), and 47TucW which was identified by the 
discovery (Edmonds et al. 2002) of its optical counterpart 
with photometric periodicity and phase identical to the 
radio values, are clearly detected (though G and I, 
with 0.12$\,^{\prime\prime}$ separation, are not resolved) in this deep Chandra image. 
Several MSPs, particularly C and T, were marginal detections in the 
original ACIS-I image but are now well detected. The total 
counts recorded in the summed Chandra images for each MSP range from 38--306  
over the 0.3--4\,keV band. This has allowed new studies 
(Bogdanov et al. 2004, hereafter BGH04) of 
the spectra and variability of this largest 
MSP sample in a single GC. We preview several key results from this 
work and provide some additional constraints on the X-ray spectra 
and properties of MSPs. 

\subsection{Colors, Spectra and Luminosities of MSPs in 47Tuc}

\begin{figure}
\vspace*{-0.3in}
\plotfiddle{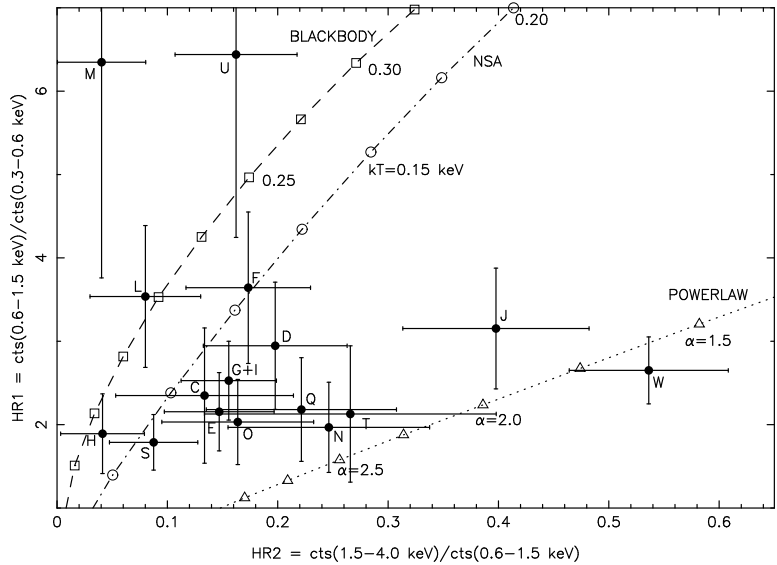}{3.5in}{-90.}{42.}{42.}{-150.}{270.}
\vspace*{-0.5in}
\caption{X-ray color-color diagram for the MSPs in 47Tuc 
(from Bogdanov et al. 2004) vs. model tracks 
folded through the detector for fixed cluster column density 
NH and blackbody emission or neutron star atmosphere 
(NSA) temperatures, or power law spectra with  
photon index values shown.}
\end{figure}

First, the {\lower0.8ex\hbox{$\buildrel >\over\sim$}}5$\times$~ 
better count statistics as well as improved 
soft response of ACIS-S (despite its low energy degradation with time) 
allow the X-ray colors and thus spectral characteristics of even 
the faintest sources to be better determined. Figure 2 shows the 
color-color diagram, derived in softer bands appropriate for ACIS-S, 
and comparison of the MSPs with tracks for blackbody (BB), neutron 
star atmosphere (NSA; Lloyd 2003) and power law (PL) models derived for 
the ACIS-S response and latest measurement 
(Gratton et al. 2003) of the cluster reddening from which 
we derive NH = $1.3 \times 10^{20}$\,cm$^{-2}$ (Heinke et al. 2004a). 
Whereas the ACIS-I color-color plot, with larger errors,  
suggested that all but MSP~J (MSP~W was not yet identified) were consistent 
with pure thermal emission (GCH02), Figure 2 suggests that most 
MSPs are a mixture of thermal and power law models, or possibly 
a two temperature thermal model and additional hard 
component, as derived by Zavlin et al. (2002) for 
the field MSP J0437$-$4715.

\begin{figure}
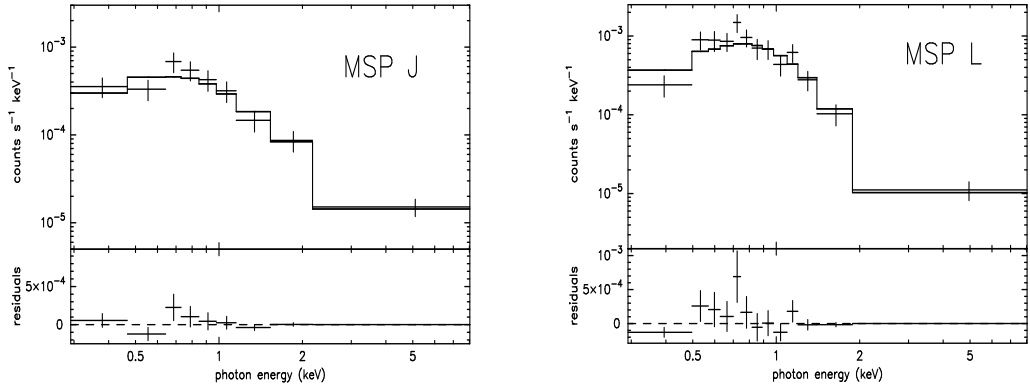


\vspace*{1.6in}
\plotfiddle{msp-J_best1.ps}{3.in}{-90.}{23.}{29.}{-200.}{360.}

\vspace*{0.3in}
\plotfiddle{msp-L_spec.ps}{3.in}{-90.}{23.}{29.}{+10.}{611.}

\vspace*{-6.in}
\caption{Spectral fits (Bogdanov et al. 2004) for two MSPs bright enough 
to clearly require hard spectral components in addition to 
a soft thermal component. MSPs O and, especially, 
W also require such a two-component model. The fits shown here are 
for a NSA + PL model but are fitted for NSA temperature 
only with PL photon index $\Gamma$ = 1 and 
NH = $1.3 \times 10^{20}$\,cm$^{-2}$~ held fixed. The derived NSA temperature 
$T$(10$^6$\,K) and radius $R$(km) values, ($T$, $R$), for J 
and L are (0.89$\pm$0.18, 
1.59$\pm$0.76) and (1.25$\pm$0.16, 1.07$\pm$0.30).}
\end{figure}

\begin{figure}
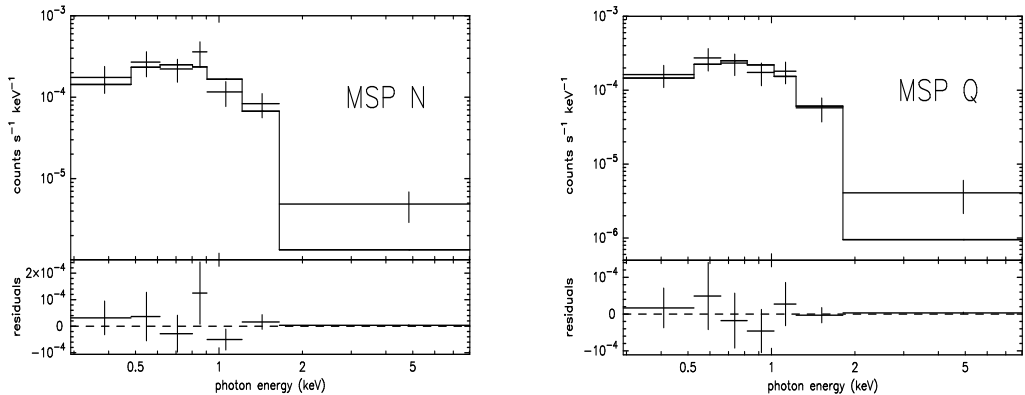


\plotfiddle{msp-N_spec.ps}{3.5in}{-90.}{23.}{29.}{-200.}{260.}
\plotfiddle{msp-Q_spec.ps}{3.5in}{-90.}{23.}{29.}{+10.}{525.}
\vspace*{-4.5in}
\caption{Spectral fits (single temperature NSA model only; from 
Bogdanov et al. (2004)) vs. data for 
fainter MSPs showing excess flux above NSA model 
in highest energy bin.}
\end{figure}


The improved statistics now allow direct spectral fits for NSA models 
(BB fits, with larger kT but smaller radii, are equally acceptable) 
and an additional PL component. Spectral fits were done 
by requiring at least 15 counts (or 10 counts for the 
faintest MSPs) per spectral bin and were thus 
limited by statistics: the highest energy bins included in the 
fit exceeded 2\,keV only for the brightest-hardest 4 MSPs: J, L, O 
and W. Spectral fits for these four were done for a NSA + PL model 
by holding the photon index fixed at 1 for the PL component. Only 
MSP~W was bright enough to fit the NSA and PL components separately,  
giving a photon index $\Gamma$ = 1.3$\pm$0.2. Spectra are shown 
in Figure 3.

For the remaining 11 MSPs (the unresolved and partly 
resolved MSP pairs, G+I and 
F+S, were each fitted as if one object), the fits were done for 
an NSA model only. However these typically showed an excess 
above the soft NSA fit at energies above 2\,keV, 
as shown in Figure 4 for MSPs N and Q. 
Thus we now conclude that although the soft-thermal emission is 
dominant, it is likely that most of the 47Tuc MSPs also contain 
an underlying harder component. It is interesting then to 
examine these thermal and hard components separately.

\subsection{Spectral properties vs. MSP properties}

The distribution of thermal temperatures, 
for an assumed neutron star atmosphere 
(NSA) model (we use the Lloyd (2003) model), is surprisingly 
narrow as shown in Figure 5a. All 
of the MSPs have their thermal emission fit by NSA models with 
mean temperatures (in units of 10$^6$ K) $T_{6} = 1.18\pm{0.22}$. 
In fact {\it all} of the MSPs except U, with $T_{6}$ = 1.83, 
have their polar cap temperatures within the narrow range of 
0.9--$1.3 \times 10^6$\,K. This presumably reflects a nearly 
constant heating rate, despite the {\lower0.8ex\hbox{$\buildrel >\over\sim$}}10$\times$ 
range of MSP ages. 

\begin{figure}
\plottwo{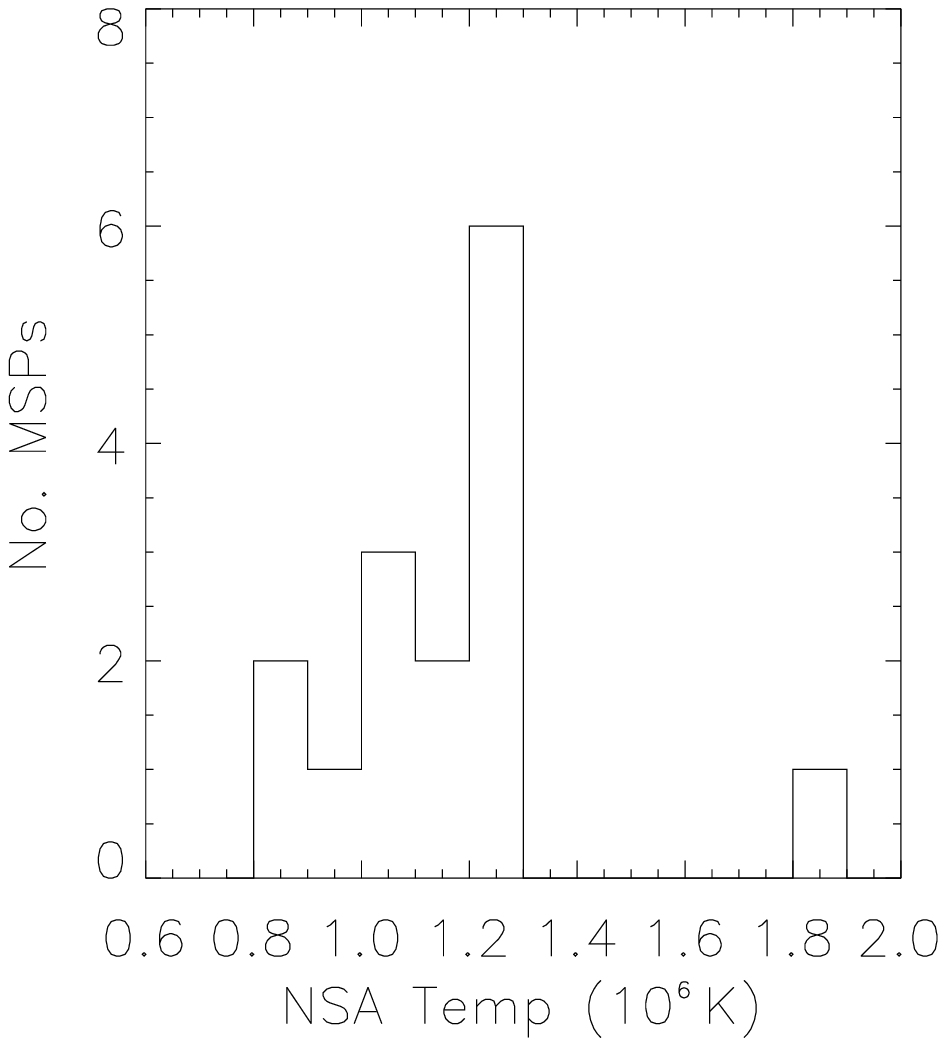}{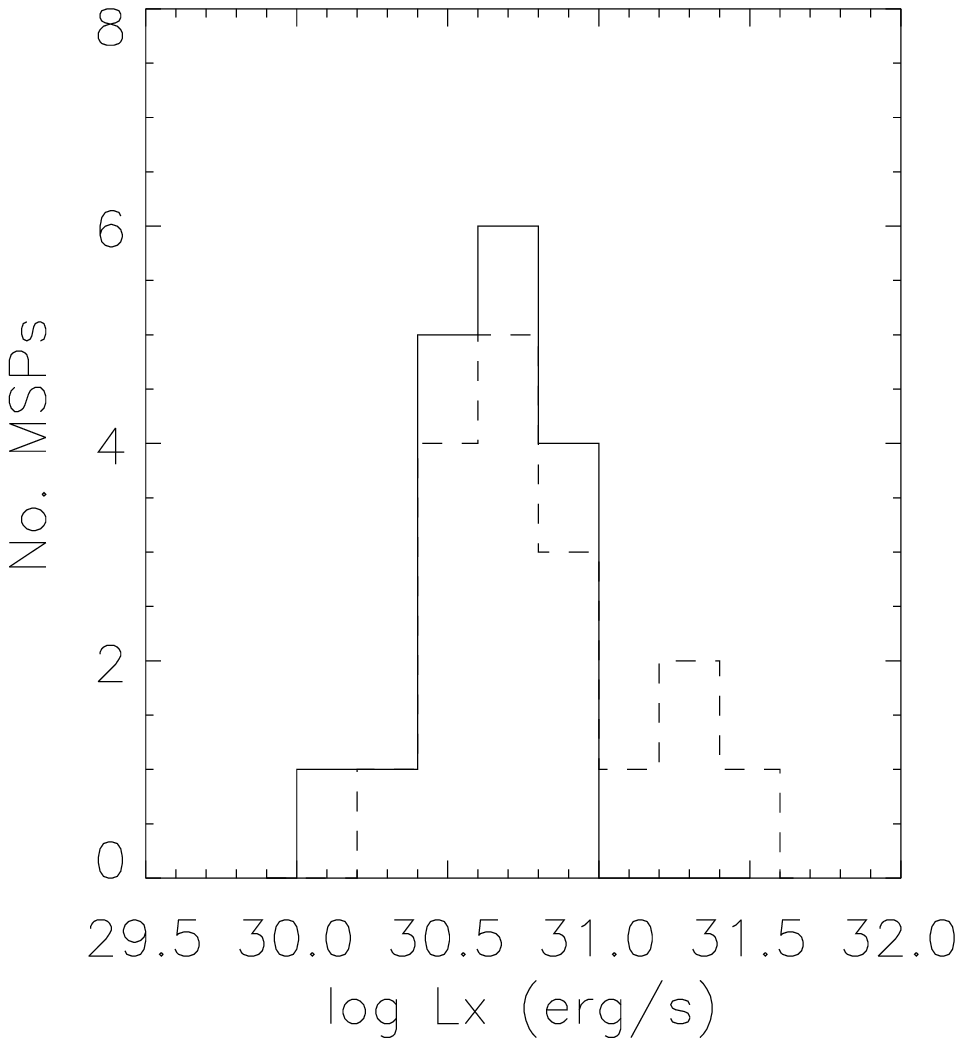}
\caption{a) Left, temperature distribution for 
NSA model fits to MSPs in 47Tuc; 
b) Right, thermal luminosity distribution (solid histogram) derived 
from $R$, $T$ values from NSA model fits vs. X-ray luminosity derived 
from total flux in 0.01--8\,keV band (dashed histogram; see text). 
The non-thermal (PL?) luminosity may be estimated from the 
difference between the total and thermal components.} 
\end{figure}

The fitted polar cap radii are also relatively narrowly distributed, 
with mean value $R$ = 0.91$\pm{0.46}$\,km, and -- given the NSA model -- is 
larger than the BB radii discussed in GCH02 as evidence the $B$ fields 
may be multipolar. The corresponding thermal 
emission luminosity, derived from a Stefan-Boltzmann application of 
the NSA-derived $R$, $T$ values, is shown in Figure 5b. 
We have assumed the thermal emission is from two polar caps, each 
of radius $R$ (both $R$ and $T$ are evaluated at the NS surface), and have 
approximated their combined luminosity as $2 \pi R^2 \sigma T ^4$, 
where $\sigma$ is the Stefan-Boltzmann constant. This is an 
approximate upper limit since it ignores the unknown projection 
effects of the polar caps due to the $B$ field inclination, and 
also ignores the exact effects of gravitational bending near the NS. 
While exact photon bending corrections are not known, they 
nevertheless ensure that both caps are seen at least partly. The 
thermal luminosity distribution (solid curve) has mean value 
log $L_{\rm thermal} = 30.64\pm0.25$ whereas the luminosity derived 
from integrating the total flux in the 0.01--8\,keV band 
(dashed histogram) has mean log $L_{\rm tot} = 30.80\pm0.32$. 
The latter luminosity was derived from the detected counts 
over this band for the two-component model (NSA + PL) spectral 
fits derived over the 0.3--8\,keV band and then integrated 
over the broad band (0.01--8\,keV) in XSPEC for an estimate 
of the total flux.

The fact that the MSP spectra can be decomposed into thermal 
and total components (Fig. 5b) suggests that the underlying 
hard component (here approximated by a PL) might be extracted 
by simple subtraction. If this hard component is indeed non-thermal 
and a PL, then this would allow constraints on the evolution of 
this magnetospheric emission vs. fundamental MSP properties such 
as magnetic field or age. In Figure 6 we show the possible dependence 
of the hard luminosity component (i.e. the excess above the NSA 
thermal flux) on a) magnetic field at the MSP light cylinder, 
$B_{\rm lc}$ , and b) the dependence 
of this same hard luminosity of spindown age. 
The intrinsic {$\dot{P}$}~ values required for either  
$B_{\rm lc}$ or $P/2\dot{P}$ are corrected for cluster acceleration as derived 
in GCH02. The MSPs display a weak (85\% confidence level from a 
Spearman rank correlation) correlation between the PL 
luminosity, $L_{\rm PL}$,    
and magnetic field at the light cylinder, $B_{\rm lc}$, and suggest 
a possible anti-correlation between $L_{\rm PL}$ and spindown age. 
Both correlations are suggested by the data presented in GCH02 
and Bogdanov et al. (2004) but are more apparent in the 
attempted decomposition of the PL or hard luminosity presented 
here. 

\begin{figure}
\plottwo{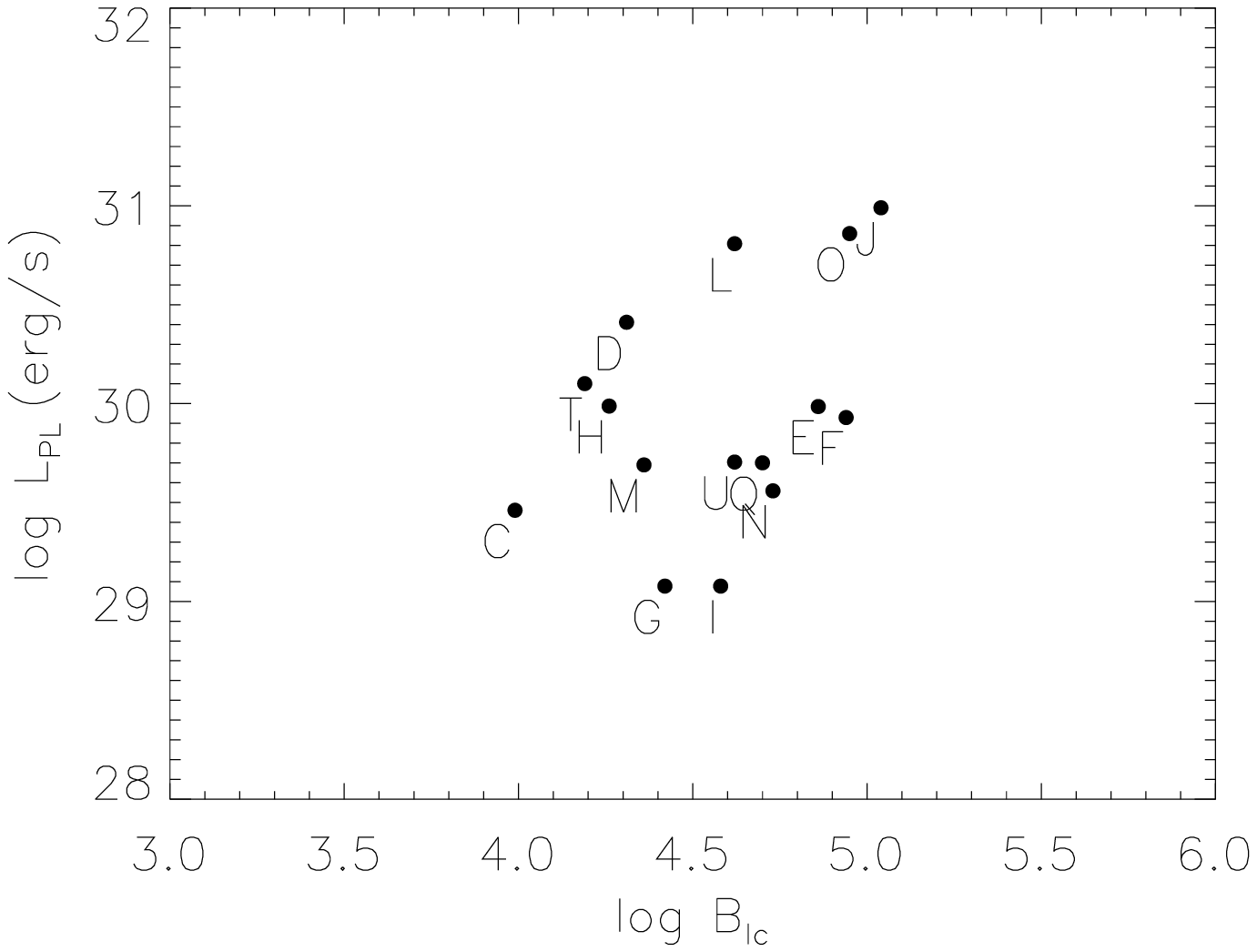}{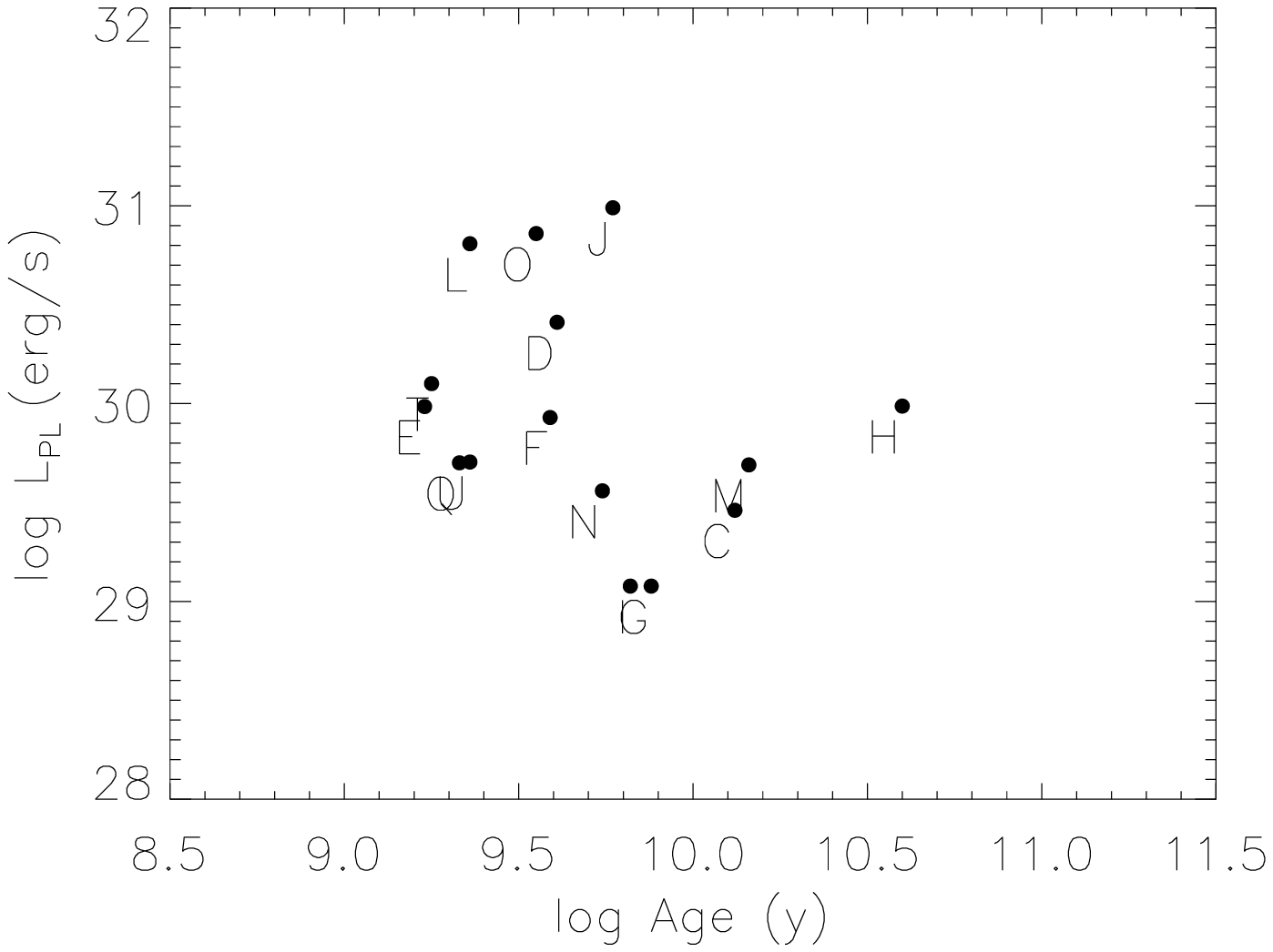}
\caption{Plot of ``PL luminosity'' (= 
$L_{0.01-8keV}$ - $L_{\rm NSA}$) for each MSP in 47Tuc 
with measured {$\dot{P}$}~ using the luminosities for each plotted in the  
distributions shown in Figure 5b) vs. a) left, magnetic field 
at the MSP light cylinder and vs. b) right, spindown age, $P/2{\dot{P}}$.}
\end{figure}
 
Note that this ``PL Luminosity'' may not be magnetospheric 
non-thermal emission, though this is plausibly at least partly the case. 
Instead, some MSPs could have their hard component due to 
their pulsar winds shock heating gas driven off from 
their binary companions, as is the case for MSP~W (BGvdB04). 
At first this seems unlikely since none of the MSPs in Figure 6 
have binary mass functions and constraints on 
inclination (e.g. eclipses) which require main sequence 
companions as is the case for MSP~W. Freire (2004, in this volume) 
discusses the distinction between the low mass binary pulsars 
(LMBPs), for which mass functions require secondaries with 
masses {\lower0.8ex\hbox{$\buildrel >\over\sim$}}0.1${M_{\odot}}$ 
and which are usually He white dwarfs, 
and  the eclipsing LMBPs (ELMBPs) such 
as MSP~W for which the main sequence (and thus massive) 
secondaries are required from their optical identifications. 
These ELMBPs are much less common than the very low mass (secondary) 
systems, the VLMBPs, with secondary masses usually 
{\lower0.8ex\hbox{$\buildrel <\over\sim$}}0.05${M_{\odot}}$. 
However, some of these VLMBPs are 
eclipsing (e.g. MSP~J, which Camilo et al. (2000) note is 
eclipsed at 70\,cm and 50\,cm, but not at 20\,cm, for about 
a quarter of its orbit) and have luminous hard 
components. A magnetospheric vs. pulsar wind shocked gas 
origin for the hard component would both likely scale with 
$B_{\rm lc}$. 

The magnetospheric vs. shocked gas origin for the ``PL''  
component could be tested by (much) more sensitive spectra 
which would also distinguish PL from a hot thermal bremsstrahlung component, 
though brems is surely ruled out by dispersion 
measure constraints, as pointed out 
by GCH02. Pulsation analysis would provide a definitive test 
since the shocked gas origin should 
not be pulsed due to the emission region having size 
(much) larger than the characteristic pulse period length scale, 
$c \cdot P \sim 10^8$\,cm. Unfortunately, detailed spectral fits 
are not possible with the present Chandra data given the limited statistics, 
and  pulsed profiles vs. energy are not possible with 
Chandra-ACIS (or Chandra-HRC). Thus we cannot prove the ``PL'' 
component is of magnetospheric origin on these old, weak-field MSPs. 

Another test can be done, however: plot the hard X-ray 
luminosity, $L_x$(PL), vs. {$\dot{E}$} and compare with the soft 
X-ray luminosity $L_x$(NSA) vs. {$\dot{E}$} (as in GCH02). If the hard 
component is magnetospheric, it will likely fall closer 
onto the roughly linear $L_x \sim {\dot{E}}$ relation shown for 
field MSPs and luminous GC-MSPs (Becker and Trumper 1999, GCH02) 
than to the $L_x \sim \dot{E}^{0.5}$ 
relation for the soft component that is likely due to polar 
cap heating (GCH02). Using the {$\dot{E}$} values for the 47Tuc MSPs 
corrected for cluster acceleration (GCH02), we plot $L_x$ vs. {$\dot{E}$}
in Figure 7 for the two components of $L_x$. The $L_x$(NSA)--{$\dot{E}$}
relation has slope $0.3\pm0.2$, which is 
marginally flatter than the slope 0.5$\pm0.2$ relation found in 
GCH02 for the dependence of the 0.5--2.5\,keV luminosity measured 
with ACIS-I for a mean BB spectrum with $kT$ = 0.22\,keV. The 
$L_x$(PL)--{$\dot{E}$} relation appears steeper, with slope $0.6\pm0.3$ from 
fitting all the points. However a slope 1 relation is also 
consistent and in fact approximately ``bisects'' 
and is parallel to the two apparent  groups of 
points (the lower of which is formed by MSPs G, I, N, U, 
Q, F, E; this grouping can also be seen in Figure 6a). 

Thus within the scatter, a linear relation between 
$\log (L_x({\rm PL}))$--{$\dot{E}$}~ is consistent with the data. 
We suspect the ``PL'' 
component lies in a broad band, bounded by the upper and 
lower groups on either side of the linear correlation line. 
As suggested above, it is possible that some of these 
on the more luminous side (e.g. J and O) are a mixture 
of magnetospheric and shocked gas components. 
It is thus interesting that MSP~O 
is similar to J in Figure 8, since it is so similar in 
its binary properties (cf. Figure 5 in Freire, in this volume). 
Conversely, the ``lower group'' of MSPs in the $\log (L_x({\rm PL}))$
vs. {$\dot{E}$}~ plot are dominated by single MSPs (F, G, N, U), 
wide binaries (E, Q) or those with extremely low mass companions  
(I, with $M_c \sim 0.01\,{M_{\odot}}$; Freire, in this volume). 

\begin{figure}
\plotone{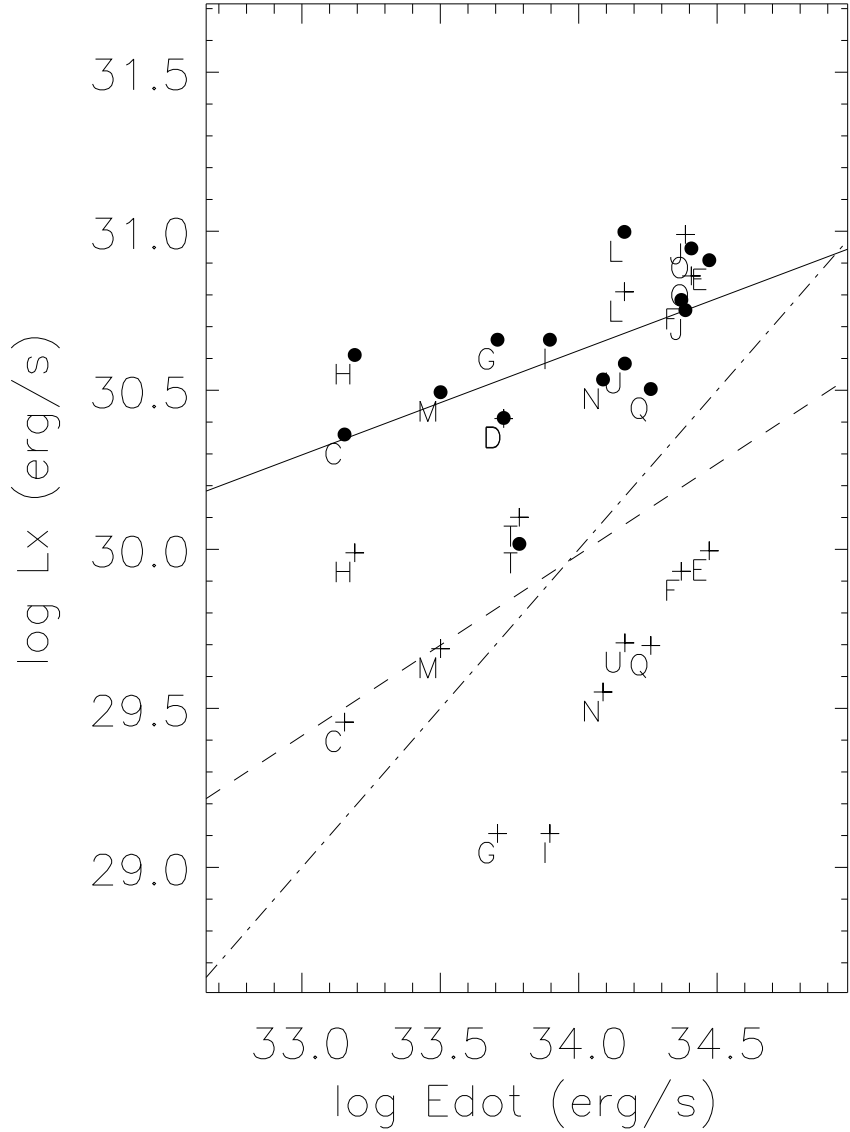}
\caption{Log-log plot for X-ray luminosity vs. 
spindown luminosity, {$\dot{E}$}~, for 
thermal (NSA) component ($\bullet$) vs. power law component (+). 
The linear regression lines shown are for the thermal luminosities  
($\log (L_x)$ = 19.5$\pm4.6$ + 0.33$\pm0.13$ $\log$({$\dot{E}$}), 
solid line), PL luminosities 
($\log(L_x)$ = 10.6$\pm11.7$ + 0.57$\pm0.35$ $\log$({$\dot{E}$}), 
dashed line), and a linear 
relation for the PL luminosities ($\log L_x$ = $-4$ + log({$\dot{E}$}~), dash-dot 
line), which bisects the upper and lower tracks of PL points and 
would be consistent with the  linear Becker and Trumper (1999) relation 
for more luminous pulsars dominated by magnetospheric emission.}
\end{figure}

\subsection{47TucW and Total MSP Populations}

We have already hinted at the special role MSP~W in 47Tuc has 
assumed. As an eclipsing (3.1h) radio MSP with a main sequence (MS) 
companion discovered with HST (Edmonds et al. 2002), it is 
similar in many respects to the 1.3d binary-eclipsing MSP 
6397A with MS (or evolved) companion 
in NCG 6397. As noted in Edmonds et al., the X-ray 
spectrum derived with the original Chandra ACIS-I observation is 
also similar: a hard spectrum (fit with a PL with photon index 
$\Gamma$ = 1.8 $\pm$0.6, with $L_x$(0.5--2.5\,keV) = $7.8 \times 10^{30}$\,erg\,s$^{-1}$ 
vs. values 1.6 $\pm0.3$ and $4 \times 10^{30}$\,erg\,s$^{-1}$ for 6397A. 

The deep Chandra observation of 47Tuc has greatly expanded our 
view of MSP~W, with more than 7$\times$ the total number of counts. 
The X-ray spectrum and variability, as well as the simultaneous 
HST/ACS-WF data and archival ACS-HRC data, are described in 
detail in BGvdB04. As included in Figure 5, we now fit W 
with an NSA model with $T = 1.03 \times 10^6$\,K and radius (at the NS) 
of 1.4km for a thermal luminosity of $L_{\rm NSA}$ = 
$7.8 \times 10^{30}$\,erg\,s$^{-1}$ and a PL component with photon index 1.3$\pm0.25$ 
and non-thermal luminosity $L_{\rm PL}$ = $2.8 \times 10^{31}$\,erg\,s$^{-1}$.  
From comparison with the qLMXB J1808.4$-$3658, 
we conclude that MSP~W  is the long-sought 
missing link between MSPs and qLMXBs in its properties and 
that it constrains the origin of the hard PL component seen 
in many qLMXBs to be due to the shocked gas due to an underlying pulsar 
wind, as recently suggested by Campana et al. (2004) and references 
therein. 

Here we simply comment on the implications for finding an 
X-ray population of hard (``PL emission'') MSPs, like W, 
for the total population of MSPs in 47Tuc or in other clusters. 
Freire (in this volume) notes that there are now 6 such eclipsing 
MSPs known in GCs (including the three with optically identified 
MS like companions: 47TucW, 6397A and PSR B1718$-$19 in 
NGC 6349; and 3 additional likely candidates in other 
GCs) out of some 80 total MSPs known in GCs. This of course 
is based on a radio-selected sample (all MSPs were originally 
discovered as radio objects). However these objects 
as well as the EVLMBPs (e.g. J) are all subject to eclipses 
and variable absorption in the radio. We note that 47TucW 
was only detected for 4h (Camilo et al. 2000), thus  
precluding a timing measurement for its apparent {$\dot{P}$}~ and 
therefore our including W with a derived $B_{lc}$ or age in 
Figure 6 or {$\dot{E}$}~ in Figure 7. Our Chandra spectra now 
also indicate that at least two of these (6397A and 47TucW) 
are similar in their PL-component X-ray 
spectra, which  are likely also due to pulsar wind shocked gas 
rather than magnetospheric emission in hard X-rays. 
Thus the Chandra deep images can detect these objects 
regardless of their radio detectability and inclination; MSP~W is dominated 
by hard emission that would not be eclipsed if the system 
were at inclination $i$\,{\lower0.8ex\hbox{$\buildrel <\over\sim$}}40$\,^{\circ}$  
but in fact is eclipsed for part of 
the binary orbit since it likely arises from shocked gas near the 
L1 point (BGvdB04). 

This suggests that some 
of the hard source population in 47Tuc 
and other GCs could be 
additional MSPs that are permanently ``eclipsed'' in radio
by the winds driven from their secondary companions by 
pulsar wind shock heating. Thus the $\sim$8\% fraction of 
ELMBPs inferred by Freire (in this volume) must be a lower limit; 
all those that are permanently eclipsed in radio are not 
counted. These hard X-ray sources would be variable (on 
binary phase timescales) and could appear as 
blue objects due to pulsar wind-heated stellar companions 
or possibly heated accretion streams, prevented 
from accreting by pulsar winds. We note that heating 
effects (and moderately blue companions) are only expected 
for the shortest orbital period systems like W; those 
which have second-exchange captured companions in to 
{\lower0.8ex\hbox{$\buildrel >\over\sim$}}6--8h 
periods (like 6397A, with 1.3d orbital day period) would likely 
not show heating effects (though H$\alpha$~ may still be detected). 
As hard sources with otherwise main sequence type optical 
counterparts, they could then be confused with active binaries 
or BY Dra systems, which is precisely what happened with 
our initial identification (Taylor et al. 2001) 
of the optical counterpart of 
6397A before the discovery (D'Amico et al. 2001) 
of the MSP in NGC 6397 and our identification (Grindlay et al. 2001b) 
of it as a Chandra source. The total number of MSPs 
in 47Tuc must then include some of these and be larger than 
the revised estimate of $\sim$30--60 MSPs and $\sim$400 NSs 
derived by Heinke et al. (2004a) 
in part from the 8\% estimate of Freire (in this volume).

\section{Are the MSPs and qLMXBs in 47Tuc Anisotropic?}

\begin{figure}
\plotone{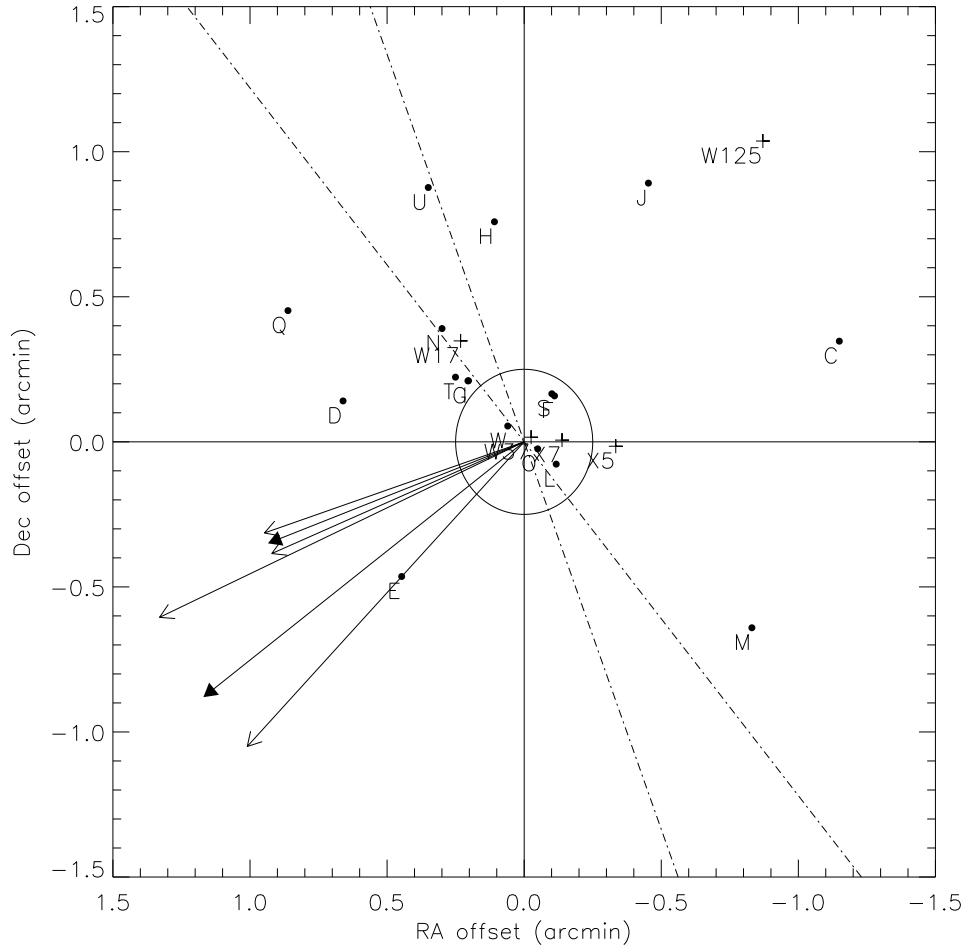}
\caption{Offsets from the cluster center (de Marchi et al. 1996) 
of all 17 MSPs ($\bullet$) with known positions and the 5 qLMXBs (+) in 
47Tuc compared with the proper motion (PM) vectors  
and $\pm1\sigma$ uncertainties $250\pm2 \,^{\circ}$ and 
$233^{+12}_{-9}\,^{\circ}$ of the cluster 
as reported by Anderson and King (2003) and 
Odenkirchen et al. (1997), respectively (we show both for comparison  
since the more accurate value is derived as an offset 
from the SMC proper motion which may still be subject to systematic 
uncertainties, though its quoted errors were included). The core radius of 
the MSPs and qLMXBs in 47Tuc, $r_c = 15\,^{\prime\prime}$~ as 
derived by GCH02, is shown as the circle. The dot-dashed lines  
are perpendicular to the PM vectors and are similarly uncertain. 
The Odenkirchen PM vector suggests the MSPs and qLMXBs may be trailing 
the cluster (7 of 22 are leading); the Anderson-King PM vector suggests 
they may be offset perpendicular to the cluster motion (7 of 22 are 
"below" the PM vector). The binomial probabilities for either are 
about 3\% and so only marginally significant (see text). }

\end{figure}

We note that the spatial distribution of the 15 MSPs in 47Tuc 
with timing positions originally 
reported by Freire et al. (2001) 
as well as that of 47TucS (Freire 2001) and now also 
47TucW (Edmonds et al. 2002) are suggestive of anisotropy: all 
but 4 (E, O, L and M) are on the north side of the 
cluster center, over a range of position angle (PA) bisectors 
through the cluster center. Interestingly, all 5
of the qLMXBs in 47Tuc (Heinke et al. 2004b) are on the same (N-NW) side 
of a bisector through the cluster center as the MSPs. 
In Figure 8 we show the positions of the 17 MSPs and 5 qLMXBs with 
precise positions in 47Tuc. The binomial probability of only 4 
(MSPs D, E, L and M) of 22  
objects that are themselves likely drawn from the same 
parent distribution (LMXBs) being on the SE side of the cluster 
center is $4.2 \times 10^{-4}$; this increases to $2.2 \times 10^{-3}$ 
for 5 objects (over a wider range of PAs). 

The significance of this result is thus 
only $\sim 3 \sigma$ and of course less if all possible ``bisectors" 
are considered. However, the possible anisotropy direction 
may be ``preferred": the MSPs (and qLMXBs) are approximately  
on the side trailing the direction of proper motion (PM) of 47Tuc, 
which is towards PA = $233^{+12}_{-9}$ $^{\circ}$~,  
as derived by Odenkirchen et al. (1997) by referencing to Hipparcos data. 
The a-posteriori binomial probability of the 7 MSPs (and no qLMXBs), 
namely MSPs Q, D, E, T, G/I and W being ``below'' the bisector 
line through the cluster center and perpendicular to the PM vector 
(double-dot-dash drawn in Figure 8) is 0.026. On the other hand, the 
more accurate PM vector at PA = $250 \pm2$ $^{\circ}$~ reported by 
Anderson and King (2003) does not suggest this but instead has a similar 
number (7) of MSPs (E, O, L, M) and qLMXBs (X7 and X5) offset to one side 
of the PM vector. This PM determination, while statistically more accurate, 
is measured relative to the SMC for which the value reported by Irwin (1999) 
contributes the dominant error and which may still contain systematic effects.    
Thus, overall, the anisotropy may be associated with the cluster proper motion 
at the $\sim2\sigma$ level.  

Regardless of any association with the cluster PM, if the 
MSP-qLMXB distribution is anisotropic, it is unlikely that selection 
effects could introduce azimuthal 
anisotropy in the Chandra source distribution (indeed 
the overall X-ray source distribution in Figure 1 is isotropic), 
and no dispersion measure (DM) gradients are present in the DM values 
given by Freire et al. (2001) that could give rise to a lack of 
radio detections of MSPs in the SE quadrant. 
In contrast, the comparable numbers of Chandra sources optically 
identified by Edmonds et al. (2003a,b) with CVs (22) and 
active binaries (29), which are  
chromospherically active main sequence binaries, appear to be  
isotropic in the deep HST images used for their identification.

Why should the MSPs/qLMXBs, predominantly, 
be affected by the cluster motion if the trailing or perpendicular 
associations are significant? 
We note that if the MSP/qLMXB offsets are real, they would not be expected 
to show up in the present PM measures for the MSPs (four are given 
in Freire et al. (2001)) vs. the cluster PM since the difference must 
be negligible for the MSPs to remain bound to the cluster. 
As the oldest compact binary population, with (typically) degenerate  
secondaries, the MSPs (and qLMXBs) may have accumulated the largest net recoil 
deflections by weak encounters with soft main sequence binaries in the 
galactic halo and disk. An MSP (or qLMXB) will have weak encounters 
with wide field binaries (with semi-major axes 
typically $a_f$ = 100AU) in the disk and galactic halo,  
which are (much) too soft to survive in the cluster core, 
at a rate $R_f \sim n_{f} \sigma_{f} v_f$. Adopting a weak 
encounter impact parameter as $b_f \sim 100 a_f$, which would 
typically Rutherford scatter a cluster MSP to induce a $\sim$1\% 
velocity perturbation on the typical $\sim$10 km\,s$^{-1}$ 
velocity of the MSP in the cluster, a
field binary density $n_f \sim 10^{-3}$\,pc$^{-3}$, 
and cluster velocity relative to the field $v_f \sim100$\,km\,s$^{-1}$,  
the time between such encounters is $\tau_f$ =$1/R_f \sim 10^{8-9}$y.  
This is comparable to the weak encounter time for an MSP  
with cluster binaries in the core of 47Tuc, 
which as hard binaries have $a_c$\,{\lower0.8ex\hbox{$\buildrel <\over\sim$}}1AU, 
$n_c \sim 10 ^3$ pc$^{-3}$ 
(for an assumed 1\% binary fraction for such wide binaries in 47Tuc), 
and relative velocity in the cluster $v_c \sim10$\,km s$^{-1}$. 
These weak encounters will be isotropic in the cluster frame for 
cluster binary perturbers but anisotropic for field binaries, which 
will impart (small) recoil momentum transfer to the MSPs and qLMXBs.  
Since the maximum transfer is at closest approach, it would 
seem the recoil velocities would be preferentially in the 
perpendicular direction. Note that such recoils would 
occur preferentially when the compact binaries are near apastron 
in their largely radial cluster orbits (where they spend the most time), 
so that relatively small $\Delta v$ values imparted to 
the binary can have the largest effect. The length of 
the PM arrows are drawn in Figure 8 for the two different PM values 
and show the approximate motion of the center of mass of the 
cluster and MSPs over 10$^4$ years. The comparable offsets of the MSPs, if 
acquired over their $\sim 10^{9}$\,year lifetimes, thus 
represent a net drift velocity of only $\sim 10^{-2}$ km\,s$^{-1}$, 
or comparable to the induced velocity perturbation by scattering off 
field binaries. 

Single cluster stars have smaller cross sections than binaries for 
scattering off field binaries, and without the internal degree 
of freedom provided by the binding energy of a 
cluster binary, scattering of single stars are likely perturbed  
more isotropically by their encounters. Cluster CVs and ABs have 
binary evolution lifetimes generally (much) shorter than MSPs and may also  
absorb their recoils in tidal distortion of their non-degenerate 
secondaries (absent for all the known MSPs in 47Tuc except W). 
Detailed simulations and a full diffusion analysis are of course needed 
to test whether disk and halo binary encounters can scatter the 
orbits of compact binaries in a cluster preferentially and not 
alter (significantly) their radial distribution (GCH02), which is 
consistent with an isothermal King profile with core radius $r_c = 15\,^{\prime\prime}$
(cf. Fig. 8) as expected for $\sim$1.5${M_{\odot}}$  objects in dynamical equilibrium 
with predominantly 0.7${M_{\odot}}$ stars in the core. 

Clearly both a larger sample of MSPs and/or qLMXBs is needed to 
test the reality of this (marginal) anisotropy as well 
as speculative interpretation. Based on X-ray luminosity and spectral 
colors and absence of variability or CV/AB 
optical counterparts, additional 
MSP candidates in the original ACIS-I data have been given by GCH02 and 
Edmonds et al. (2003a) but have been reduced  by Heinke et al. (2004a) 
to just 4 or 5 in the ACIS-S data: W5, W28, W34, W142 and 
possibly W6. All of these 
additional MSP candidates are again perpendicular (and on one side) 
of the Anderson-King PM vector 
but only two (W34 and W142) are on the ``trailing'' side. This would 
support the picture of preferentially perpendicular (to the PM vector) 
scattered velocity perturbations, though it is puzzling these are 
not symmetrically distributed about both 
sides rather than just the north side.   
In addition, of course, there are at least 5 radio 
MSPs in 47Tuc still not located with 
radio timing, and as noted above, Heinke et al. (2004a) 
estimate there are an additional $\sim$10--40 additional MSPs likely 
present in 47Tuc (likely a lower limit; see above). 
Thus MSP/qLMXB possible anisotropy and diffusion effects will be testable.

\section{Conclusions}
MSPs in GCs remain a treasure to be mined further with high 
spatial resolution 
X-ray imaging as well as high throughput spectroscopy. They are low 
luminosity objects which must be resolved, cleanly, from comparably 
bright CVs and ABs. It is clear that Chandra and future higher throughput telescopes, 
with comparable or (ideally) better angular resolution ($< 1\,^{\prime\prime}$) 
are needed. With Chandra, very long observation times are required to 
make significant new progress on MSPs. Accordingly, we have concentrated here on 
X-ray spectra derived from our 300ksec observation of 47Tuc and have 
shown that they are generally all consistent with a combination of 
soft thermal emission from their polar caps plus a harder PL component. 
The thermal temperatures (derived for a NSA model) are in a 
surprisingly narrow range: $1.1 \pm 0.2 \times 10^6$ K. 
The ``PL component" can only be fitted as such in the four brightest 
MSPs; for the others we can only say there is an excess of flux above 
the thermal fits at 
{\lower0.8ex\hbox{$\buildrel >\over\sim$}}2\,keV which is consistent with a PL origin. 

We note that this hard X-ray component could be either magnetospheric or 
pulsar-wind shocked gas or both. Shocked gas is certainly 
the case for the brightest ``hard" MSP in 47Tuc, 
MSP~W (BGvdB04). MSPs, generally, may have 
all three components -- polar cap, magnetospheric and pulsar wind-induced, 
with the latter most prominent for the doubly-exchanged systems with main 
sequence companions (47TucW and 6397A).

It is interesting that the luminosities separately 
derived here for the thermal (NSA or BB) 
and PL components have correlations with {$\dot{E}$} consistent with log-log 
slope 0.5 and 1, respectively. The PL component decreases in luminosity 
with MSP age and field at the light cylinder, both of course due to 
variations in {$\dot{P}$}. We note again that all {$\dot{P}$} values are the same 
intrinsic values corrected for the MSP acceleration in the cluster as 
described in GCH02.

We have called attention to a puzzling, though marginally significant, 
apparent anisotropy of the MSPs and qLMXBs in 47Tuc: this may (or may not) 
be associated with the cluster proper motion. Weak 
encounters the oldest cluster compact binaries, the MSPs and 
qLMXBs -- NSs with (typically) degenerate binary 
companions -- with soft binaries in the 
field (both halo and disk) may provide an explanation, but detailed 
simulations are needed and, of course, more MSP/qLMXB precise positions 
are required to confirm the effect.

X-ray pulse timing and pulse-phased spectroscopy is the obvious new 
domain which would provide crucial tests. The soft thermal component emission 
should show up as broad quasi-sinusoidal emission, which can test the 
polar cap geometry (and presence of multipoles; cf GCH02). The magnetospheric 
component is beamed (at least partly) and will 
generally produce narrower pulse 
profiles dominated by PL emission; whereas the PL component from the 
pulsar wind and shocked gas from the binary companion will be unpulsed. 
It is unfortunate that the only pulsar timing capability on Chandra 
is the HRC-S with its limited sensitivity above 2\,keV, high internal 
background, lack of any spectral sensitivity 
and limited response above $\sim$2\,keV. 
Nevertheless, the 800\,ksec HRC-S observation to be conducted in Cycle 6 
by Rutledge et al. will make an important start on measuring these 
predicted components (though the sensitivity for the PL component 
will be limited) and may enable Chandra identification of the 
remaining 5 radio MSPs by detecting their soft pulsations from among the 
ACIS-S sources.  
 
\vspace*{0.5in}
I thank Craig Heinke and Slavko Bogdanov for many discussions and for 
preparing Figures 1 and 2--4, respectively, and Andrew Lyne for discussions 
about the puzzling possible MSP anisotropy. This work was 
supported in part by 
NASA grants GO2-3059A and  HST-GO-09281.01-A.

\end{document}